\documentclass[prl,showpacs,floatfix,twocolumn]{revtex4}
\usepackage{graphicx}
\usepackage{dcolumn}

\begin{document}

\newcommand*{\cm}{cm$^{-1}$\,}
\newcommand*{\nco}{Na$_x$CoO$_2$\,}

%
\title{Infrared spectroscopy of the charge ordering transition in Na$_{0.5}$CoO$_2$}
%
%

\author{N. L. Wang}
\email{nlwang@aphy.iphy.ac.cn}%
\affiliation{Institute of Physics and Center for Condensed Matter
Physics, Chinese Academy of Sciences, P.~O.~Box 603, Beijing
100080, P.~R.~China}
\author{Dong Wu}
\author{G. Li}
\affiliation{Institute of Physics and Center for Condensed Matter
Physics, Chinese Academy of Sciences, P.~O.~Box 603, Beijing
100080, P.~R.~China}
\author{X. H. Chen}
\author{C. H. Wang}
\author{X. G. Luo}
\affiliation{Structure Research Laboratory, University of Science and Technology of China, Hefei 230026, P. R. China}
%
%
%
\begin{abstract}
We report infrared spectra of a Na$_{0.5}$CoO$_2$ single crystal
which exhibits a sharp metal-insulator transition near 50 K due to
the formation of charge ordering. In comparison with x=0.7 and
0.85 compounds, we found that the spectral weight associated with
the conducting carriers at high temperature increases
systematically with decreasing Na contents. The charge ordering
transition only affects the optical spectra below 1000 cm$^{-1}$.
A hump near 800 cm$^{-1}$ develops below 100 K, which is
accompanied by the appearance of new lattice modes as well as the
strong anti-resonance feature of phonon spectra. At lower
temperature $T_{co}$, an optical gap develops at the magnitude of
2$\Delta\approx3.5k_BT_{co}$, evidencing an insulating charge
density wave ground state. Our experimental results and analysis
unequivocally point towards the importance of charge ordering
instability and strong electron-phonon interaction in \nco system.
\end{abstract}


\maketitle
The recent discovery of superconductivity at 5K in hydrated sodium
cobaltate, Na$_x$CoO$_2\cdot yH_2O$,\cite{Takada} has stimulated
many studies on correlated electrons in a two-dimensional
triangular lattice. The precursor host compound of this
superconductor is unhydrated Na$_x$CoO$_2$. It consists of
alternate stacks of Na and CoO$_2$ layers with edge sharing
CoO$_6$ octahedra. The physical properties of Na$_x$CoO$_2$ depend
strongly on the Na concentration. Recent transport, magnetic and
structural studies on single crystals of \nco for $0.3<x<0.75$
revealed a crossover from an unusual Curie-Weiss metal near x=0.7
to a paramagnetic metal for x near 0.3. The composition that
separates the two metallic regimes at higher and lower Na
concentrations, Na$_{0.5}$CoO$_2$, undergoes a transition into an
insulating state at 53 K, accompanied by a giant increase of Hall
coefficient. Electron diffraction studies revealed the presence of
an orthorhombic symmetry of superlattice in Na$_{0.5}$CoO$_2$,
which was attributed to Na ordering.\cite{Foo}. Neutron powder
diffraction measurements further revealed an ordering of the Na
ions into zigzag chains along one crystallographic direction,
which decorates the chains of Co ions with different amounts of
charges.\cite{Huang}

Charge ordering in \nco is a subject of great interest. It was
suggested to be a major instability in the narrow conduction band
of $CoO_2$ layer, in addition to the
superconductivity.\cite{Baskaran} Charge ordering at commensurate
fillings x=1/4 and 1/3 were studied in detail and regarded as a
competitor for the superconductivity observed in the range of
$1/4\leq x\leq1/3$ in hydrated \nco.\cite{Baskaran,Kunes,Lee}
Possible charge ordered states at other commensurate fillings
x=1/2, 2/3 and 3/4 were also proposed.\cite{Baskaran,Kunes,Lee}
Those charge ordered states are believed to be easily frustrated
by the random potential from the neighboring Na layers, resulting
in a glassy phase, which is likely responsible for the anomalous
metallic behavior. Experimentally, although NMR measurements point
towards possible charge orderings in \nco for $0.5\leq
x\leq0.75$,\cite{Cavilano,Mukhamedshin} an unambiguous charge
ordering state with localized electrons was only observed for
x=0.5, as we have mentioned above.

Infrared spectroscopy is a powerful tool to probe the charge
excitations of an electronic system. Infrared investigations on
metallic \nco with different Na concentration have been reported
by several groups.\cite{Lupi,Wang,Bernhard,Caimi} In this work, we
present the in-plane optical measurements at different
temperatures on Na$_{0.5}$CoO$_2$ single crystals, focusing on the
evolution of the electronic states across the charge ordering
transition. In comparison with x=0.7 and 0.85 compounds, we found
that the spectral weight associated with the conducting carriers
at high temperature increases systematically with decreasing Na
contents. The charge ordering transition only affects the
conductivity spectrum below 1000 cm$^{-1}$. A broad hump near 800
cm$^{-1}$ develops below 100 K, with its intensity further
enhanced at lower temperature. Upon entering the charge ordering
state, a sharp suppression of the spectral weight is seen near 250
cm$^{-1}$, indicating the opening of a charge gap.

High-quality Na$_{0.5}$CoO$_2$ single crystals with size around
2mm$\times$2mm were obtained by flux growing method and chemical
deintercalation of Na in solutions of I2 dissolved acetonitrile.
Detailed preparation and characterization of the samples will be
published elsewhere.\cite{Chen} The near-normal incident
reflectance spectra were measured on the freshly cleaved surface
by a Bruker 66v/S spectrometer in the frequency range from 40 \cm
to 29000 \cm, as described in our earlier report.\cite{Wang}
Standard Kramers-Kronig transformations were employed to derive
the frequency-dependent conductivity spectra.

\begin{figure}[t]
\centerline{\includegraphics[width=3.3in]{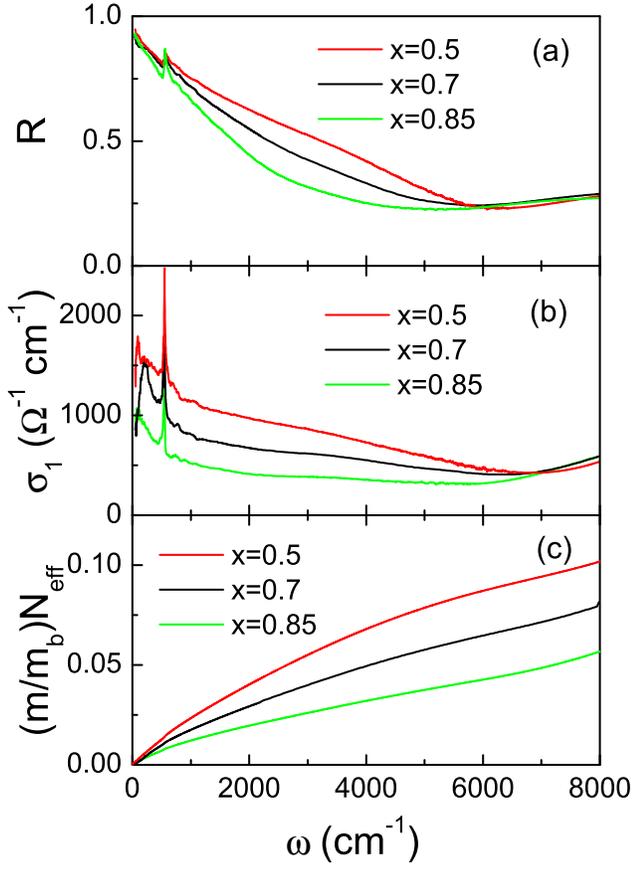}}%
\vspace*{0cm}%
\caption{Frequency dependences of the in-plane reflectance (a),
conductivity (b) and effective density of carriers per Co ion (c)
for \nco with different Na concentrations at room temperature.}%
\label{fig2}
\end{figure}

Fig. 1(a) and (b) show the room temperature in-plane reflectance
and conductivity spectra for Na$_{0.5}$CoO$_2$ together with two
metallic \nco crystals with higher Na concentrations x=0.7 and
0.85. Here, the spectra of x=0.7 crystal were taken from our
earlier measurement.\cite{Wang} The crystal of x=0.85 was grown by
a floating zone optical image furnace. The transport and magnetic
properties of this crystal was presented elsewhere.\cite{Luo} The
effective density of carriers per Co ion contributed to
conductivity below $\omega$ can be obtained by the partial sum
rule
\begin{equation}
  \frac{m}{m_b}N_{eff}(\omega)=\frac{2mV_{cell}}{\pi{e^{2}}N}\int_{0}^{\omega}\sigma(\omega')d\omega',
\end{equation}
where m is the free-electron mass, m$_b$ the averaged
high-frequency optical or band mass, $V_{cell}$ a unit cell
volume, N the number of Co ions per unit volume. Fig. 1(c)
displays $N_{eff}$ as a function of frequency for the three
samples. It shows clearly that the effective conducting carriers
increases with decreasing Na contents, even though x=0.5 compound
becomes insulating at low temperature. $N_{eff}$ can be related to
an equivalent plasma frequency, after choosing a proper
high-frequency limit $\omega_c$, via the relationship
$\omega_p^2=4\pi{e^{2}}N_{eff}(\omega_c)/m_b(V_{cell}/N)=8\int_{0}^{\omega_c}\sigma(\omega')d\omega'$.
Choosing $\omega_c\approx$6000 \cm, a frequency where $R(\omega)$
reaches its minimum but below the interband transition, we get the
overall plasma frequency $\omega_p\approx$1.4$\times10^4$ \cm,
1.2$\times10^4$ \cm and 1.0$\times10^4$ \cm for x=0.5, 0.7 and
0.85 compounds, respectively. The results strongly suggest that
metallic \nco should be considered as a doped band insulator with
the hole concentration of (1-x), rather than as a doped Mott
insulator with the electron concentration of x. Since the infrared
spectra with higher Na concentrations 0.58$\leq x\leq0.82$ have
been reported, we shall focus our attention on the x=0.5 sample in
the rest of this paper.

\begin{figure}[t]
\centerline{\includegraphics[width=3.3in]{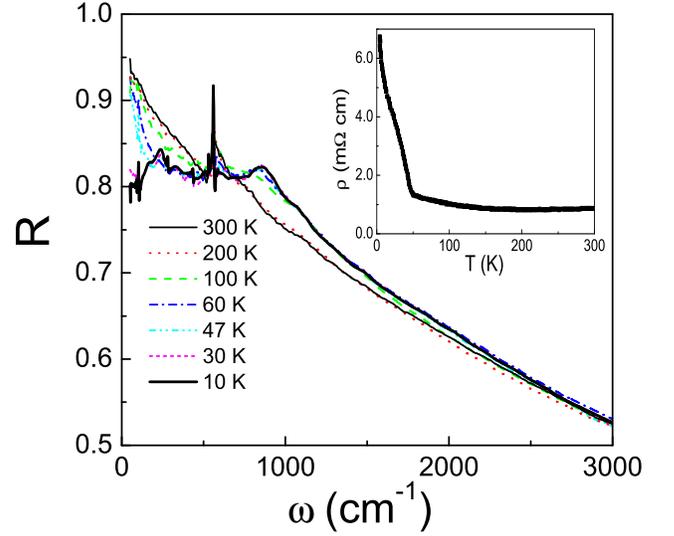}}%
\vspace*{0cm}%
\caption{Frequency dependences of the in-plane reflectance spectra
for Na$_{0.5}$CoO$_2$ at different temperatures.
The inset shows the curve of dc resistivity vs temperature of the Na$_{0.5}$CoO$_2$ crystal.}%
\label{fig2}
\end{figure}

The temperature dependent reflectance of Na$_{0.5}$CoO$_2$ crystal
is shown in Fig. 2. The inset shows the temperature dependent
in-plane dc resistivity $\rho_{ab}$ determined by four-contact
method, which is very similar to the result by Foo.\cite{Foo}
$\rho_{ab}$ has a weak temperature dependence at high temperature
but increases sharply near $T_{co}\approx50$ K, the characteristic
temperature for the charge ordering transition. In accord with
nonmetallic dc resistivity behavior with a negative slope, we
found that the low-frequency reflectance decreases with decreasing
temperature. However, the reflectance in the mid-infrared region
increases with decreasing temperature. Upon entering the charge
ordering state (T$<T_{co}$), $R(\omega)$ below 250 \cm was further
suppressed, leading to an energy gap in the extracted conductivity
spectra.

Fig. 3 shows the low frequency conductivity spectra of
Na$_{0.5}$CoO$_2$ at different temperatures. The room temperature
spectrum is very similar to the measurement result of x=0.7
compound, except for a relatively higher magnitude of conductivity
curve due to higher conducting carrier density. The low frequency
drop of $\sigma_1(\omega)$ is believed to have a common origin as
in x=0.7 compound, an issue being addressed in our earlier
work.\cite{Wang} As the temperature decreases from 300 K to 200 K,
the spectral change is very samll. However, at lower temperatures,
a number of striking features manifest in $\sigma_1(\omega)$
spectra. First of all, a sharp suppression of the conductivity
spectra below the charge ordering transition temperature $T_{co}$
is seen near 250 cm$^{-1}$, being indicative of the opening of a
charge gap. The magnitude of the gap, 2$\Delta$, defined as an
onset energy of the steeply rising part of $\sigma_1(\omega)$ is
roughly 125 \cm at our lowest measurement temperature, leading to
the value of 2$\Delta/k_BT_{co}$=3.5, which is in good agreement
with the predicted value for a mean-field charge density wave
(CDW) transition. A peak could be seen just above the gap edge,
which is also a predicted feature of CDW transition, and has been
observed in many CDW materials. The observation provides
convincing evidence for the formation of an insulating charge
density wave ground state in the ordered phase.

\begin{figure}[t]
\centerline{\includegraphics[width=3.3in]{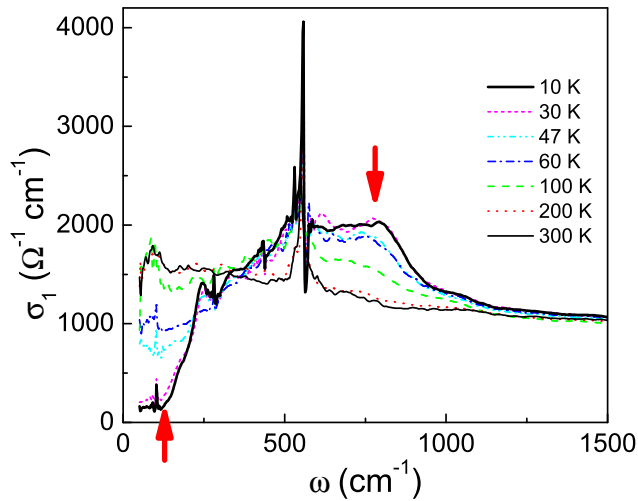}}%
\vspace*{0.3cm}%
\caption{The in-plane optical conductivity spectra of
Na$_{0.5}$CoO$_2$. The arrow at low frequency side indicates the gap position,
the arrow at higher frequency side indicates a hump, which is enhanced at lower temperature.}%
\label{fig3}
\end{figure}

Secondly, a broad hump near 800 cm$^{-1}$ develops at low
temperature. The feature is already evident at 100 K, a
temperature much higher than $T_{co}$, but becomes further
enhanced at lower temperature. The hump is a novel phenomenon for
x=0.5 compound. It is absent for other x concentrations of \nco
system.\cite{Lupi,Wang,Bernhard,Caimi} The hump position sets up
an energy scale where the charge carriers become frozen or
bounded. At present, its origin is not clear. To some extent, the
hump is related to the charge ordering since its spectral weight
comes from the missing area in the low frequency part. On this
account, its emergence above $T_{co}$ may be explained as due to
fluctuated charge ordering. However, the charge ordering itself
does not necessarily cause such a hump feature. In many other
charge ordering compounds, no such hump was observed. A favorable
candidate for the hump is that it is caused by a polaronic
characteristic of charge carriers due to the enhanced
electron-phonon interaction at low temperature. The temperature
dependence of the hump is in good agreement with the expected
behavior of polarons. Further support for this possibility comes
from phonon spectra, as we shall discuss below.

Thirdly, dramatic change appears in phonon modes. At high
temperatures (above 200 K), only two infrared active phonons are
present in the spectra: a stronger one at 551 \cm and a weaker one
at a bit lower frequency 530 \cm. The two phonons are similar to
the results seen in x=0.7 crystal at high temperature,\cite{Wang}
both being close to the frequency of a hard E$_{1u}$ mode as
predicted by symmetry analysis.\cite{Li} However, below 100 K,
three additional phonon modes at 102, 282, and 435 \cm could be
seen clearly. The appearance of the new phonon modes suggests the
change of the structure. Since the modes could already be seen at
100 K, being correlated with the above hump feature, the
structural instability occurs at much higher temperature than the
metal-insulator transition temperature. Electron diffraction
measurements by Huang et al. on Na$_{0.5}$CoO$_2$ revealed extra
diffraction spots at 100 K in comparison with diffraction pattern
at room temperature, being indicative of a structure distortion at
low temperature.\cite{Huang} We believe that the appearance of new
phonon modes is correlated with this structure distortion, which
could be ascribed to the ordering of Na ions. On the contrary, the
metal-insulator transition at 50 K is associated with the charge
ordering transition of Co ions. In addition to the appearance of
new modes, another striking observation is that the phonons at low
temperature exhibit extremely strong antiresonance feature or Fano
lineshape at the region where the electronic background is high.
By contrary, such lineshape is almost completely absent at high
temperature. Such antiresonance feature unambiguously indicates a
strong electron-phonon coupling in Na$_{0.5}$CoO$_2$. As a result,
it may cause the formation of localized bounded states of charge
carriers, or small polarons.

The above experimental observation and analysis lead us to arrive
at following picture for the evolution of the charge dynamics with
temperature in Na$_{0.5}$CoO$_2$ compound. The charge carriers at
low temperature should be regarded as bounded small polarons due
the strong electron-phonon coupling. They start to form at around
100 K, leading to the hump feature in the conductivity spectra. At
lower temperature $T_{co}$, a CDW order, perhaps a "small polaron
CDW", is further formed, which is accompanied by a gap opening in
the charge excitation spectrum.

We now discuss the implications of the optical data. From the
comparison of x=0.5 with x=0.7 and 0.85 compounds, we have shown
that the effective conducting carriers increases with decreasing
Na contents, even though x=0.5 compound becomes insulating at low
temperature. However, the available ARPES experiments indicated
that the "Fermi surfaces" of Na$_{0.7}$CoO$_2$
compound\cite{Hasan} is larger than that of Na$_{0.5}$CoO$_2$
compound\cite{Valla}. In this case, the Luttinger theorem for
enclosed volume of Fermi surface is apparently violated in doped
\nco system. We believe that those seemingly contradicted results
pose a strong constraint on a theory. One theoretical approach,
which is capable to explain the anomalous metallic properties at
low temperature and unusual "Fermi surfaces" seen in ARPES for
\nco, was developed by Baskaran.\cite{Baskaran} In this theory,
the underlying charge ordering at commensurate filling is a key
assumption. However, due to the random potential from the
neighboring Na layers and the strong commensurability effects of
the triangular lattice, the charge ordered states would be easily
frustrated. Baskaran suggested that the metallic state of \nco
system is a homogeneous quantum molten state of these ordered
states, which he referred to as a quantum charge liquid.

Although charge ordering in \nco system was suggested to be a
major instability in the narrow conduction band of $CoO_2$ layer,
static charge ordering with localized electrons was only observed
for x=0.5 compound at low temperature. Our infrared data show that
even for the x=0.5 compound, the charge dynamics is affected only
at very low frequencies (roughly below 1000 \cm). We emphasize
that this energy scale is much lower than those found for many
other systems. For example, in quasi-one-dimensional charge
ordering systems BaIrO$_3$ or organic systems like (TMTSF)$_2$X
salts, the size of optical gap (2$\Delta$) is around
9$k_BT_c$.\cite{Cao,Jacobson} In La$_{1-x}$Ca$_x$MnO$_3$ system,
the ratio of the 2$\Delta/k_BT_{co}$ could be as large as 30 for
x=0.5.\cite{Kim} The very small energy scale may explain why
charge ordering is easily be destroyed and not observed in other
commensurate filling x in \nco.

To conclude, we have measured the infrared spectra of
Na$_{0.5}$CoO$_2$ which is known to have static charge ordering at
low temperature. In comparison with x=0.7 and 0.85 compounds, we
found that the spectral weight associated with the conducting
carriers at high temperature increases systematically with
decreasing Na contents. The charge ordering transition only
affects the conductivity spectrum below 1000 cm$^{-1}$. A broad
hump near 800 cm$^{-1}$ develops below 100 K, which is correlated
with the appearance of new lattice modes as well as the strong
anti-resonance feature of phonon spectra. At lower temperature
$T_{co}$, a CDW order, perhaps a "small polaron CDW", is further
formed, which is accompanied by the development of a gap in the
charge excitation spectrum. Our work highlights the importance of
charge ordering and strong electron-phonon interaction in \nco
system.

We acknowledge helpful discussions with J. L. Luo, T. Xiang, and
G. M. Zhang. This work is supported by National Science Foundation
of China (No. 10025418, 10374109), the Knowledge Innovation
Project of Chinese Academy of Sciences.

%
%

\end{document}